%% file: draft.tex
\documentclass[aps,prd,twocolumn,showpacs,amsmath,amssymb]{revtex4-1}
\usepackage{amsmath} \usepackage{graphicx} \usepackage{subfigure}
\usepackage{epstopdf} \usepackage{color} \usepackage{multirow}
\usepackage{setspace} \usepackage{overpic} \usepackage{amssymb}
\usepackage{ulem}
\usepackage{siunitx}

\usepackage[bookmarksnumbered, pdfstartview=FitH,colorlinks,urlcolor=blue, citecolor=blue,linkcolor=blue] {hyperref}
\usepackage{lineno}
\usepackage{bm}
\usepackage{rotating}
\usepackage{xcolor}
\usepackage{makecell}
\usepackage{mathtext}
\usepackage{mathrsfs}
\usepackage[utf8]{inputenc}
\usepackage{threeparttable}

\let\oldequation\equation
\let\oldendequation\endequation
\renewenvironment{equation}
 {\linenomathNonumbers\oldequation}
 {\oldendequation\endlinenomath}

\include{def-com-new}

\begin{document}

\title{\boldmath Observation of the decay $h_{c}\to3(\pi^{+}\pi^{-})\pi^{0}$ }

\input{authorlist_2023-11-13.tex}
\date{\today}

\begin{abstract}
 Based on $(2712.4\pm14.1)\times10^{6}$ $\psi(3686)$ events collected with the BESIII detector, we  study the decays $h_{c}\to3(\pi^{+}\pi^{-})\pi^{0}$, $h_{c}\to2(\pi^{+}\pi^{-})\omega$, $h_{c}\to2(\pi^{+}\pi^{-})\pi^{0}\eta$, $h_{c}\to2(\pi^{+}\pi^{-})\eta$, and $h_{c}\to\ppb$ via $\psi(3686)\to\pi^{0}h_{c}$. The decay channel $h_{c}\to3(\pi^{+}\pi^{-})\pi^{0}$ is observed for the first time, and its branching fraction is determined to be $\left( {9.28\pm 1.14 \pm 0.77} \right) \times {10^{ - 3}}$, where the first uncertainty is statistical and the second is systematic. In addition, first evidence is found for the modes  $h_{c} \to 2(\pi^{+}\pi^{-})\pi^{0}\eta$ and $h_{c}\to2(\pi^{+}\pi^{-})\omega$ with significances of 4.8$\sigma$ and 4.7$\sigma$, and their branching fractions are determined to be $(7.55\pm1.51\pm0.77)\times10^{-3}$ and $\left( {4.00 \pm 0.86 \pm 0.35}\right) \times {10^{ - 3}}$, respectively. No significant signals of $h_c\to 2(\pi^+\pi^-)\eta$ and $h_{c}\to p\bar{p}$ are observed, and the upper limits of the branching fractions of these decays are determined to be $<6.19\times10^{-4}$ and $<4.40\times10^{-5}$ at the 90\% confidence level, respectively.
\end{abstract}

\maketitle
\section{Introduction}
Studies of charmonium decays play an important role in understanding the structure of quantum chromodynamics (QCD). Despite the success of QCD in describing many aspects of the strong interaction, some applications in charmonium decay mechanism remain challenging and inconsistencies between experimental results and theoretical predictions have been reported~\cite{theory_review_a}.

Since the discovery of the spin-singlet charmonium state $h_{c}(^1P_{1})$ in 2005~\cite{hc_a,hc_b}, many theoretical and experimental efforts have been made to understand its properties. Kuang predicted the branching fraction (BF)  $\mathcal{B}(h_c \to $light hadrons) to be $(8.8\pm0.8)\%$ and $(48\pm7)\%$ with the perturbative QCD(PQCD) and non-relativistic QCD (NRQCD) models~\cite{Kuang:2002hz}, respectively. Godfrey made a prediction of $\mathcal{B}(h_c \to ggg)=57\%$ based on QCD calculations~\cite{Godfrey:2002rp}. The significant differences between the results of these calculations arise from the use of different computational methods based on various theoretical models and the consideration of different effects. Recently BESIII reported the BF of the electric dipole (E1) transition to be $\mathcal{B}(h_c \to \gamma\eta_c) = (57.66^{+3.62}_{-3.50}\pm 0.58)\%$~\cite{newBF_Marco2022}, which is reasonably close to some theoretical predictions~\cite{Kuang:2002hz,Godfrey:2002rp}. This implies that nearly half of $h_c$ decays are non-E1 modes. However, the sum of the BFs of the known $h_{c}$ non-E1 decays is only about 4\%~\cite{pdg2022,hc_zhul}, indicating that the search for new $h_c$ decay modes is well motivated.

In massless QCD models, the $h_{c}(^1P_{1})$ decay into $p\bar{p}$ is forbidden due to the helicity selection rule~\cite{heliciity_forbidden_hc_ppb}. However, the observation of the decays $\eta_{c}/\chi_{c0}\to p\bar{p}$~\cite{pdg2022} and the observation of $h_{c}$ formation in the $p\bar{p}$ annihilation~\cite{Andreotti:2005vu} indicate substantial contributions from the effects of finite masses. These observations have stimulated theoretical calculations for the decay $h_{c}\to p\bar{p}$, arising from which a large BF of the order of $10^{-3}$ is suggested~\cite{lch_ref5, lch_ref6}. However, searches until now have not observed a significant signal~\cite{lch_prd}. Further experimental study of this decay is highly desirable to  understand better the underlying dynamical mechanism in $h_c\to p\bar p$ decay.

In this paper, we report searches for the undiscovered decay channels of $h_c\to3(\pi^{+}\pi^{-})\pi^{0}$, $h_c\to2(\pp)\omega$, $h_c\to2(\pi^{+}\pi^{-})\pi^{0}\eta$, $h_c\to2(\pi^{+}\pi^{-})\eta$, and $h_c\to p\bar{p}$ via $\psipp\to\pi^0 h_c$, performed through the analysis of $(2712.4\pm14.1)\times10^{6}$ $\psi(3686)$ events~\cite{psip_num_0912} collected with the BESIII detector in 2009, 2012 and 2021.
\section{BESIII DETECTOR AND MONTE CARLO SIMULATION}
\label{sec:BES}

The BESIII detector~\cite{Ablikim:2009aa} records $e^+ e^-$ collisions provided by the BEPCII storage ring~\cite{CXYu_bes3} in the center-of-mass energy range from 2.0 to 4.95~GeV,
with a peak luminosity of $1 \times 10^{33}\;\text{cm}^{-2}\text{s}^{-1}$
achieved at $\sqrt{s} = 3.77\;\text{GeV}$. BESIII has collected large data samples in this energy region~\cite{Ablikim:2019hff,EcmsMea,EventFilter}. The cylindrical core of the BESIII detector covers 93\% of the full solid angle and consists of a helium-based multilayer drift chamber (MDC), a plastic scintillator time-of-flight system (TOF), and a CsI(Tl) electromagnetic calorimeter (EMC), which are all enclosed in a superconducting solenoidal magnet providing a 1.0~T magnetic field (0.9 T in 2012). The magnet is supported by an octagonal flux-return yoke with modules of resistive plate muon counters (MUC) interleaved with steel. The charged-particle momentum resolution at 1~GeV/c is 0.5\%, and the  d$E/$d$x$ resolution is 6\% for the electrons from Bhabha scattering. The EMC measures photon energy with a resolution of 2.5\% (5\%) at 1~GeV in the barrel (end-cap) region. The time resolution of the TOF barrel part is 68~ps, while that of the end-cap part is 110~ps. The end-cap TOF system was upgraded in 2015 using multi-gap resistive plate chamber technology, providing a time resolution of 60~ps, which benefits $\sim83\%$ of the data used in this analysis~\cite{tof_a,tof_b,tof_c}.

Monte Carlo (MC) simulated data samples produced with a {\sc geant4}-based~\cite{geant4} software package, which includes the geometric description~\cite{detvis} of the BESIII detector and the detector response, are used to optimize the event selection criteria, to estimate the signal efficiency and the level of background. The simulation models the beam-energy spread and initial-state radiation in the $e^+e^-$ annihilation using the generator {\sc kkmc}~\cite{kkmc_a,kkmc_b}. The inclusive MC sample includes the production of the $\psi(3686)$ resonance, the initial-state radiation production of the $J/\psi$ meson, and the continuum processes incorporated in {\sc kkmc}. Particle decays are generated by {\sc evtgen}~\cite{evtgen_a,evtgen_b} for the known decay modes with BFs taken from the Particle Data Group  (PDG)~\cite{pdg2022} and {\sc lundcharm}~\cite{lundcharm_a,lundcharm_b} for the unknown ones. Final-state radiation from charged final-state particles is included using the {\sc photos} package~\cite{photos}.

For the exclusive MC simulation samples, the five channels of interest are generated using a phase-space (PHSP) model for each signal mode.
\section{EVENT SELECTION}
\label{sec:selection}

Charged tracks reconstructed in the MDC are required to originate from a region within 10~cm from the nominal interaction point (IP) along the symmetry axis of the MDC ($z$ axis), and within 1~cm in the perpendicular plane. The track polar angle $\theta$ measured with respect to the $z$ axis must be within the fiducial volume of the MDC, $\left| {\cos\theta } \right| < 0.93$. The tracks are assumed to be protons for $h_{c}\to p\bar{p}$ decay and pions for other decay modes. Finally, a vertex fit constraining all charged particles to originate from a common vertex is performed.

Photon candidates are reconstructed from isolated electromagnetic showers
produced in the crystals of the EMC. The deposited energy of each shower must be greater than 25 MeV in the barrel region ($|\!\cos\theta|<0.80$) or greater than 50 MeV in the end-cap region  ($0.86<|\!\cos\theta|<0.92$). To suppress electronic noise and energy depositions not associated with the event, the EMC cluster timing is required to be within 700~ns of the start time of the reconstructed event. To form a $\pi^0$ candidate, the invariant mass of the selected $\gamma\gamma$ pair is constrained to the known $\pi^{0}$  mass~\cite{pdg2022} by a one-constraint (1C) kinematic fit, and the $\chi^2_{\rm 1C}$ is required to be less than 20. To form an $\eta$ candidate, the invariant mass of the selected $\gamma\gamma$ pair is constrained to the known $\eta$ mass~\cite{pdg2022}, and the $\chi_{1\rm{C}}^{2}$ is required to be less than 200.

To further suppress background, a (4+N)C kinematic fit is performed, which constrains the 4-momentum of the final state to that of the initial system, while the masses of all $\pi^{0}$ and $\eta$ candidates (denoted by N) are constrained to their nominal masses. The kinematics of the final state tracks are updated through the above kinematic fit.
A requirement is placed on the $\chi_{(4+N)\rm{C}}^{2}$ value of the (4+N)C fit  depending on the final state (see Table~\ref{list:mass_windows}).
The value of the requirement is determined by maximizing the figure of merit (FOM), defined as $S/\sqrt {S + B}$.
Here, $S$ and $B$ are the expected numbers of signal and background events, respectively, obtained from the MC simulation.
If there is any excess of photon candidates in an event, then all combinations are considered
and the one with the smallest $\chi_{(4+N)\rm{C}}^{2}$ is kept.

To suppress contamination from decays with unexpected numbers of photons, such as $\psi(3686)\to\gamma\chi_{c2}$ with $\chi_{c2}$ decaying to the same final states as the $h_{c}$, we require $\chi^{2}_{4{\rm C},n\gamma}$ $<$ $\chi^{2}_{4{\rm C},(n-1)\gamma}$ for $h_{c}\to3(\pi^{+}\pi^{-})\pi^{0}$, $h_{c}\to2(\pi^{+}\pi^{-})\omega$, and $h_{c}\to2(\pi^{+}\pi^{-})\pi^{0}\eta$. Here $\chi_{4{\rm C},n\gamma}$ is obtained from a 4C kinematic fit including $n$ photons expected for the signal candidate, while $\chi^{2}_{4{\rm C},(n-1)\gamma}$ is determined from an additional 4C fit with one missing photon compared to the signal decay, respectively. Similarly, in the $h_{c}\to2(\pi^{+}\pi^{-})\pi^{0}\eta$ mode, the background of $\psipp\to2(\pi^{+}\pi^{-})3\pi^{0}$ is suppressed by requiring $\chi _{7{\rm C}}^2 \left( 2\pi^{0}\eta2(\pi^{+}\pi^{-})\right) < \chi _{7{\rm C}}^2\left( {{3\pi ^0}2(\pi^{+}\pi^{-})} \right)$.

The $J/\psi$-related background is vetoed by the requirements on the $\pi^{0}\pi^{0}$, $\pi^{+}\pi^{-}$,
and $\eta$ recoil-mass windows. The mass windows are listed in Table~\ref{list:mass_windows}, where
$RM$ denotes the recoiling mass, defined as $RM(X)=\sqrt{(p_{\psi(3686)}-p_{X})^{2}}$,
where $p_{\psi(3686)}$ and $p_{X}$ are the four momenta of $\psi(3686)$ and X, respectively.
For the $h_{c}\to p\bar{p}$ reconstruction the background from lepton pairs originating from $J/\psi$ decays
is suppressed by requiring $|\!\cos\theta_{p(\bar{p})}|<0.8$, where $\theta_{p(\bar{p})}$ is the polar angle
of the proton (antiproton) candidate. The energy of the  $\pi^0$ from the $h_c$ decay
is typically larger than the energy of the bachelor $\pi^0$ from
$\psi(3686)\to\pi^0 h_c$. In events with two neutral pions, that one with smaller energy is denoted $\pi^0_L$
while the other is denoted $\pi^0_H$.
The bachelor $\pi^{0}_L$ should not form any resonance in combination with other final-state particles.
Therefore, additional vetoes, summarized in Table~\ref{list:mass_windows}, are applied to suppress background from $\eta,\omega\to\pi^+\pi^-\pi^0_L$.

\begin{figure}[htbp]
\begin{center}

\begin{minipage}[t]{0.9\linewidth}
\includegraphics[width=1\textwidth]{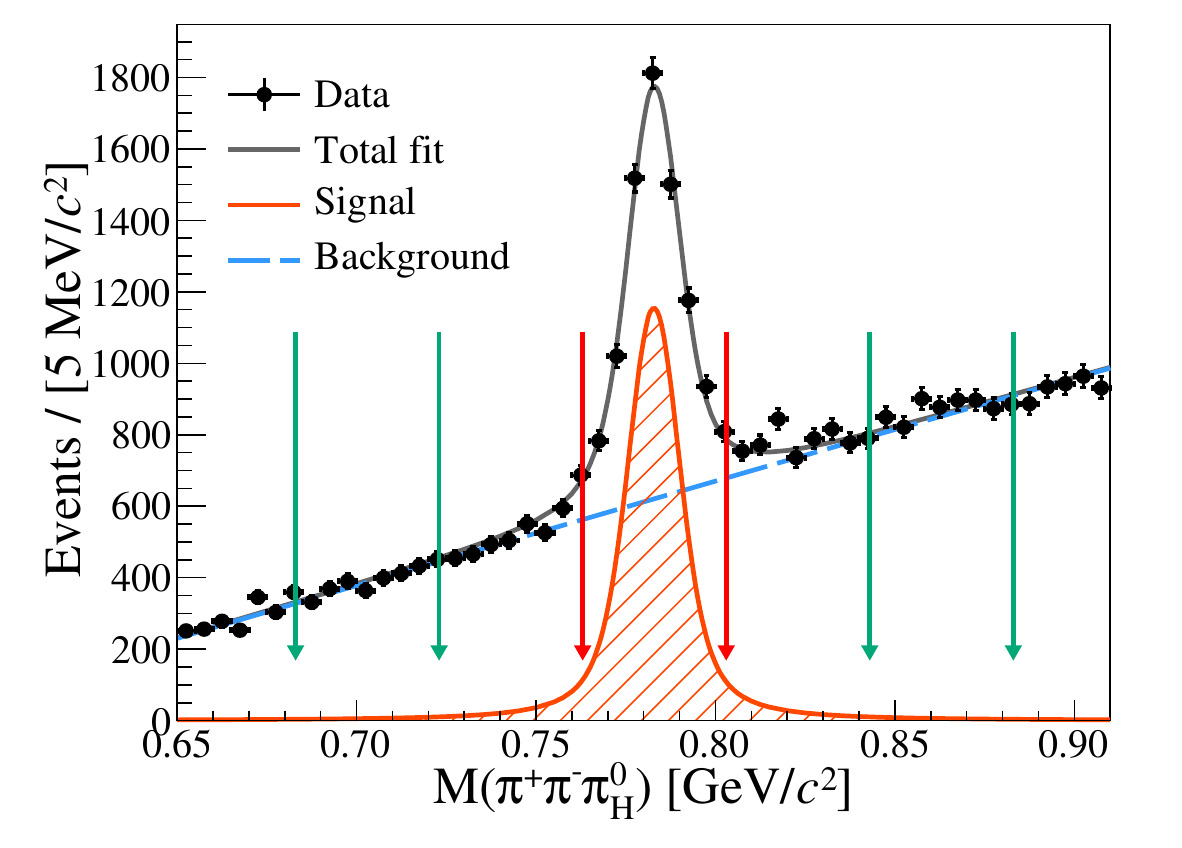}
\end{minipage}

\caption{The $M_{\pi^{+}\pi^{-}\pi_{H}^{0}}$ distribution of the accepted candidates in data. The dotted error bars represent the distribution of different $\pi^{+}\pi^{-}\pi_{H}^{0}$ combinations overlaid. The pair of red arrows shown the $\omega$ signal region, and the pair of green arrows show the $\omega$ sideband region. } \label{newFig:mass_resolution}
\end{center}
\end{figure}

In the case of candidate $h_{c}\to3(\pi^{+}\pi^{-})\pi^{0}$ decays, we examine the invariant-mass spectrum
of $\pi^{+}\pi^{-}\pi_{H}^{0}$ to check for the presence of an $\omega$ intermediate state.
The $\omega$ signal region for $h_c\to 2(\pi^+\pi^-)\omega$
is set to be $|M(\pi^{+}\pi^{-}\pi_{H}^{0})-m_{\omega}|<20$ MeV/$c^{2}$ and the $\omega$ sideband region
is set to be  $60<|M(\pi^{+}\pi^{-}\pi_{H}^{0})-m_{\omega}|<100$ MeV/$c^{2}$,
where $m_{\omega}$ is the known $\omega$  mass~\cite{pdg2022}, as shown in Fig.~\ref{newFig:mass_resolution}. All mass windows are determined to account
for the respective mass resolutions.

\begin {table}[htbp]
\begin{center}
{\caption {Requirements on $\chi_{(4+N){\rm C}}^{2}$, mass
windows and the polar angle of the proton or antiproton ($\!\cos\theta_{p(\bar{p})}$) for different signal decays.
$M$ and $m$ denote the measured invariant mass and the
known mass~\cite{pdg2022} of the indicated particle, respectively.
}
\label{list:mass_windows}}
\begin{tabular}{l c  c}
  \hline \hline

  Mode  & $\chi_{(4+N){\rm C}}^{2}$ & Veto (MeV/$c^{2}$) \\   \hline

   $3(\pi^{+}\pi^{-})\pi^{0}$   & $<35$ &$\left| {RM{{\left( {{\pi ^ + }{\pi ^ - }} \right)}} - {m_{{J \mathord{\left/ {\vphantom {J \psi }} \right.  \kern-\nulldelimiterspace} \psi }}}} \right|> 12$ \\
   &  &  $\left| {RM{{\left( {{\pi ^ 0 }{\pi ^ 0 }} \right)}} - {m_{{J \mathord{\left/ {\vphantom {J \psi }} \right.  \kern-\nulldelimiterspace} \psi }}}} \right|>30$  \\
   &  & $\left| {M\left( {{\pi ^ {+} }{\pi ^ {-} }{\pi_{L} ^{0} }} \right) - {m_{\eta} }} \right|>17$  \\    \hline

   $2(\pi^{+}\pi^{-})\pi^{0}\eta$ & $<20$  &$\left| {RM{{\left( {{\pi ^ + }{\pi ^ - }} \right)}} - {m_{{J \mathord{\left/ {\vphantom {J \psi }} \right.  \kern-\nulldelimiterspace} \psi }}}} \right|> 8$ \\

   &  & $\left| {RM{{\left( {{\pi ^ 0 }{\pi ^ 0 }} \right)}} - {m_{{J \mathord{\left/ {\vphantom {J \psi }} \right.  \kern-\nulldelimiterspace} \psi }}}} \right|>18$  \\

   &  & $\left| {RM{{\left( \eta \right)}} - {m_{{J \mathord{\left/ {\vphantom {J \psi }} \right.  \kern-\nulldelimiterspace} \psi }}}} \right|>7$  \\

   &  & $\left| {M\left( {{\pi ^ {+} }{\pi ^ {-} }{\pi_{L} ^{0} }} \right) - {m_{\eta} }} \right|>12$  \\
   &  & $\left| {M\left( {{\pi ^ {+} }{\pi ^ {-} }{\pi_{L} ^{0} }} \right) - {m_{\omega} }} \right|>26$  \\  \hline

   $2(\pi^{+}\pi^{-})\eta$ & $<25$  &$\left| {RM{{\left( {{\pi ^ + }{\pi ^ - }} \right)}} - {m_{{J \mathord{\left/ {\vphantom {J \psi }} \right.  \kern-\nulldelimiterspace} \psi }}}} \right|> 15$ \\

      &  &   $\left| {RM{{\left( \eta \right)}} - {m_{{J \mathord{\left/ {\vphantom {J \psi }} \right.  \kern-\nulldelimiterspace} \psi }}}} \right|>10$  \\

   &  &     $\left| {M\left( {{\pi ^ {+} }{\pi ^ {-} }{\pi_{L} ^{0} }} \right) - {m_{\eta} }} \right|>15$  \\
   &  &     $\left| {M\left( {{\pi ^ {+} }{\pi ^ {-} }{\pi_{L} ^{0} }} \right) - {m_{\omega} }} \right|>26$  \\  \hline

    $p\bar{p}$ & $<15$  &  ${|\!\cos\theta_{p(\bar{p})}|}<0.8$ \\

  \hline   \hline
\end{tabular}

\end{center}
\end{table}


The invariant-mass distributions for the four $h_{c}$ decay modes and the intermediate process $h_{c}\to2(\pi^{+}\pi^{-})\omega$, after all selection criteria, are shown in Fig.~\ref{fit:five_Mode_mhc}. The potential backgrounds are investigated with the inclusive $\psi(3686)$ MC events, using the event-type analysis tool TopoAna~\cite{zhouxy_topoAna}. It is found that for the $h_{c}\to2(\pi^{+}\pi^{-})\pi^{0}\eta$ selection there are peaking background contributions from $\psi(3686)\to\pi^{0}h_{c},h_{c}\to\gamma\eta_{c},\eta_{c}\to2(\pi^{+}\pi^{-})\eta$ or $\eta_{c}\to2(\pi^{+}\pi^{-}\pi^{0})$, and for the $h_{c}\to2(\pi^{+}\pi^{-})\eta$ selection there are peaking background contributions from $h_{c}\to\gamma\eta_{c},\eta_{c}\to\pi^{+}\pi^{-}\eta^{'},\eta^{'}\to\pi^{+}\pi^{-}\gamma$ or $\eta_{c}\to2(\pi^{+}\pi^{-})\pi^{0}$. Misidentification of photons as $\pi^{0}/\eta$ in these background processes can contaminate the signal region.
The size of these background contributions is estimated through MC studies and included in the fit.  The BF of the decay
 $\eta_{c}\to2(\pi^{+}\pi^{-})\pi^{0}$ is currently unknown, which is reflected in the assignment of the systematic uncertainities. Additionally, the BFs for the other mentioned background processes are taken from the PDG~\cite{pdg2022}.

To investigate contamination from continuum background, the same selection criteria are applied to the data samples collected at the center-of-mass energy of 3.65 GeV, corresponding to an integrated luminosity of 454 $\rm{pb}^{-1}$. No $h_{c}$ candidates are found. Therefore,  continuum background is considered to be at a negligible level.
\section{\label{Sec:BR_determined}Signal yields}
\begin{figure*}[htbp]

\begin{minipage}[t]{0.445\linewidth}
\includegraphics[width=1\textwidth]{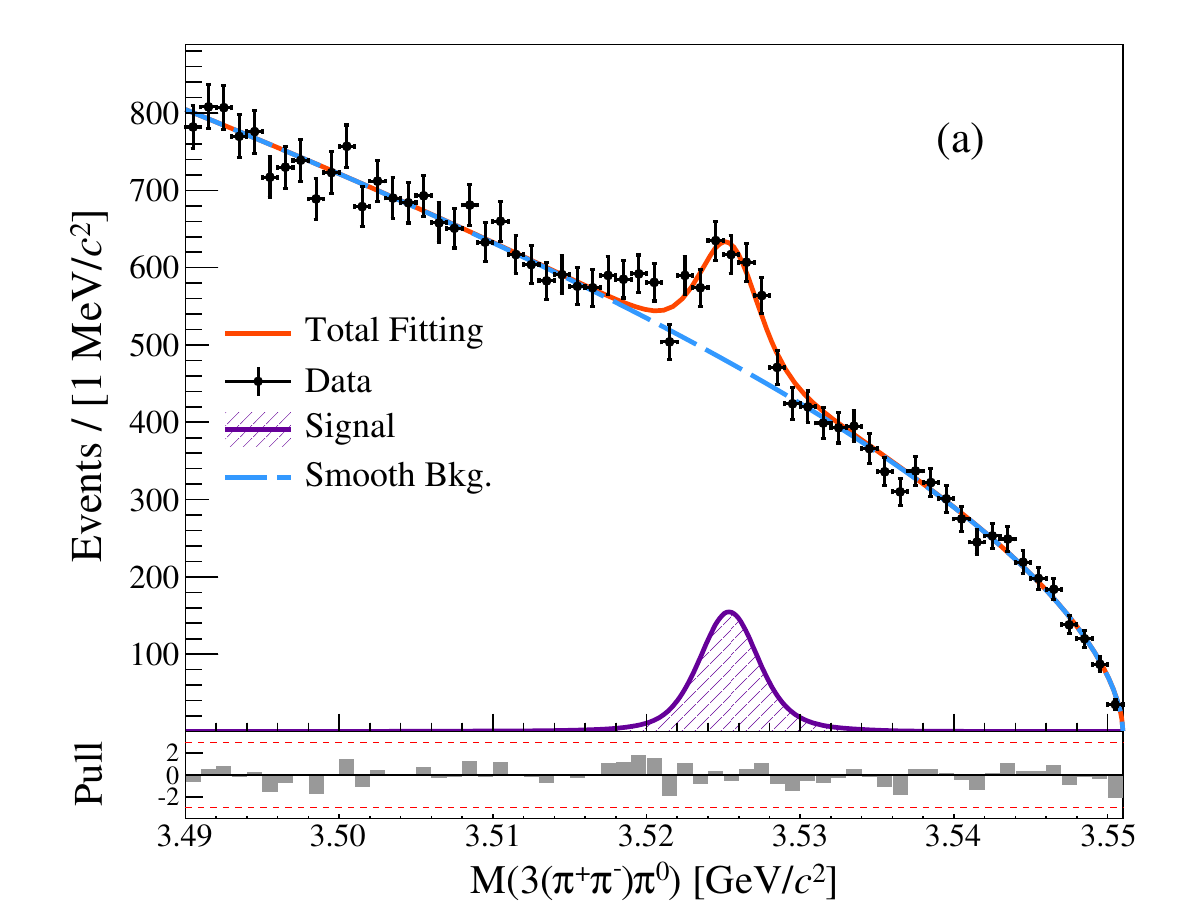}
\end{minipage}
\begin{minipage}[t]{0.445\linewidth}
\includegraphics[width=1\textwidth]{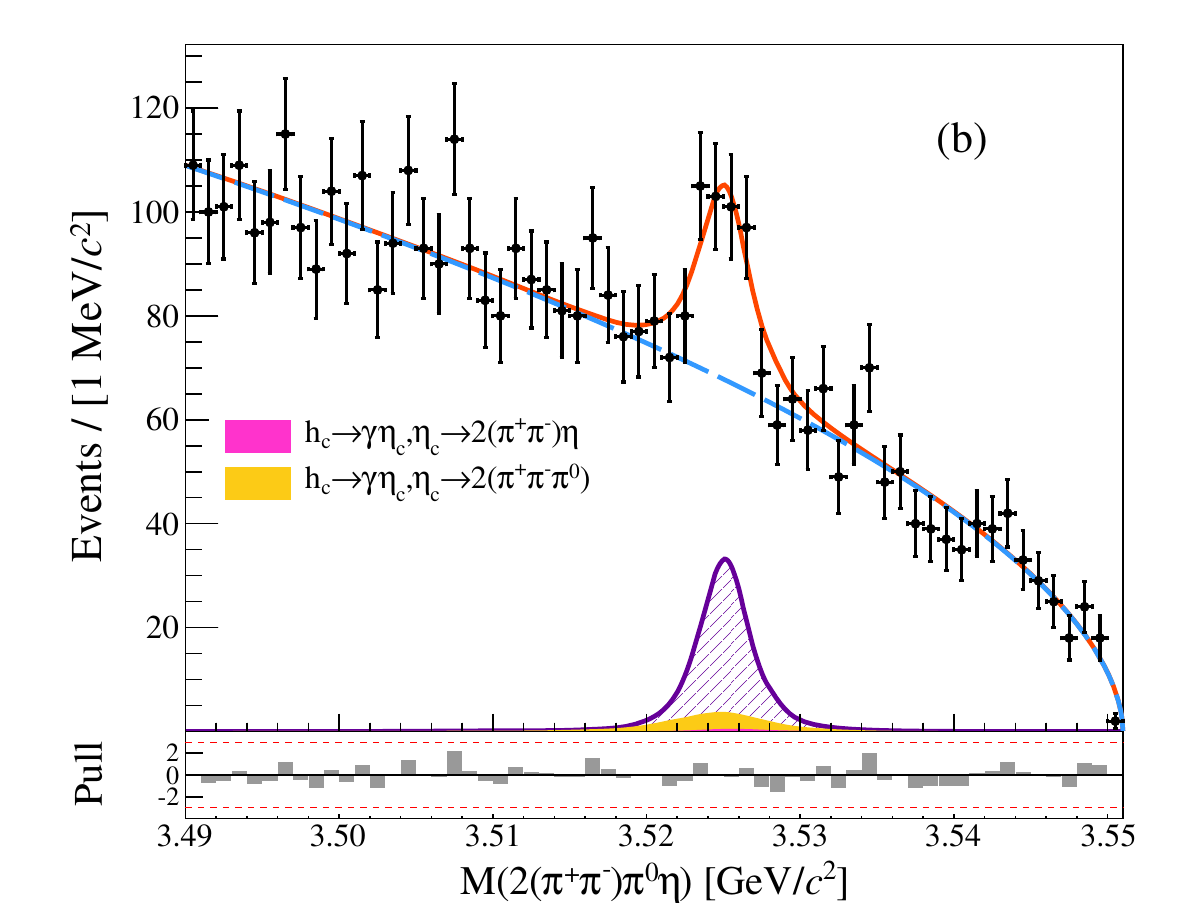}
\end{minipage}
\begin{minipage}[t]{0.445\linewidth}
\includegraphics[width=1\textwidth]{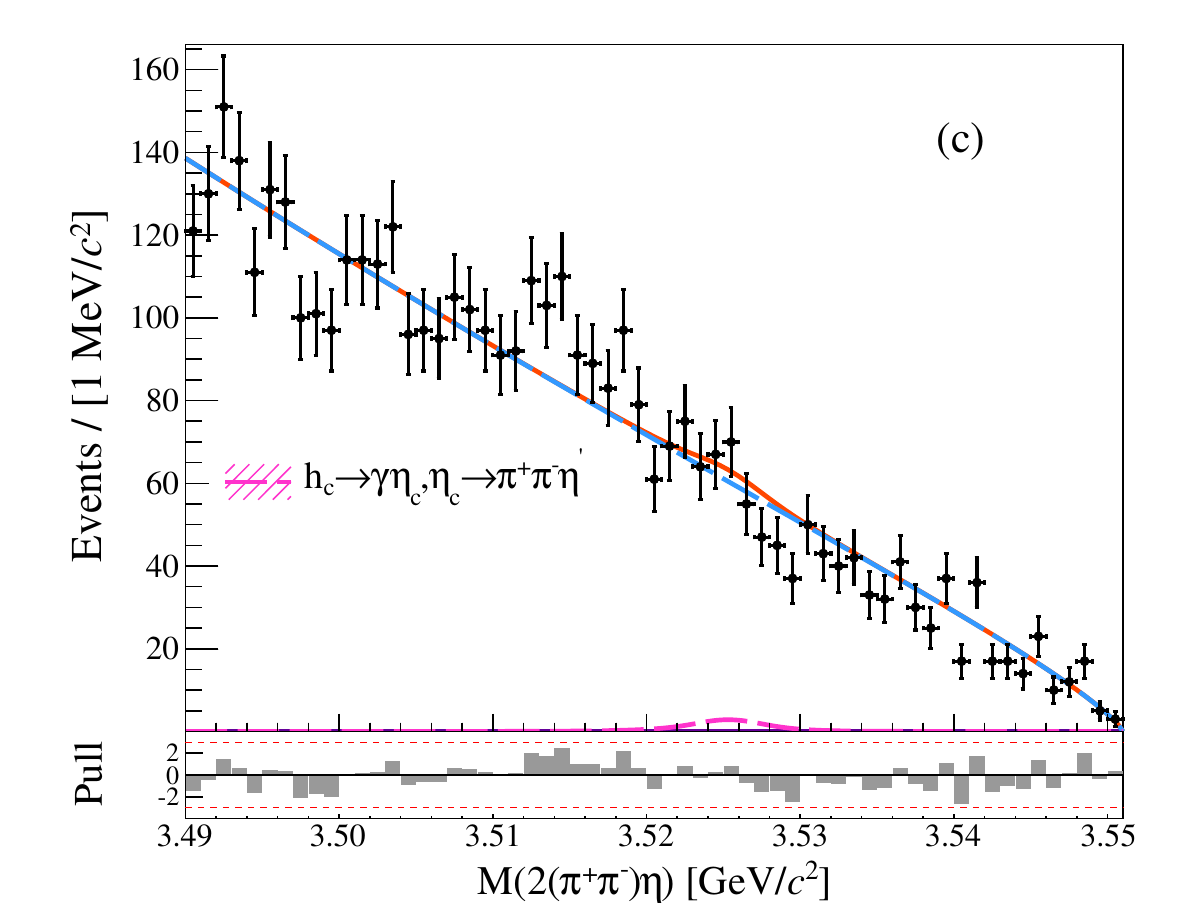}
\end{minipage}
\begin{minipage}[t]{0.445\linewidth}
\includegraphics[width=1\textwidth]{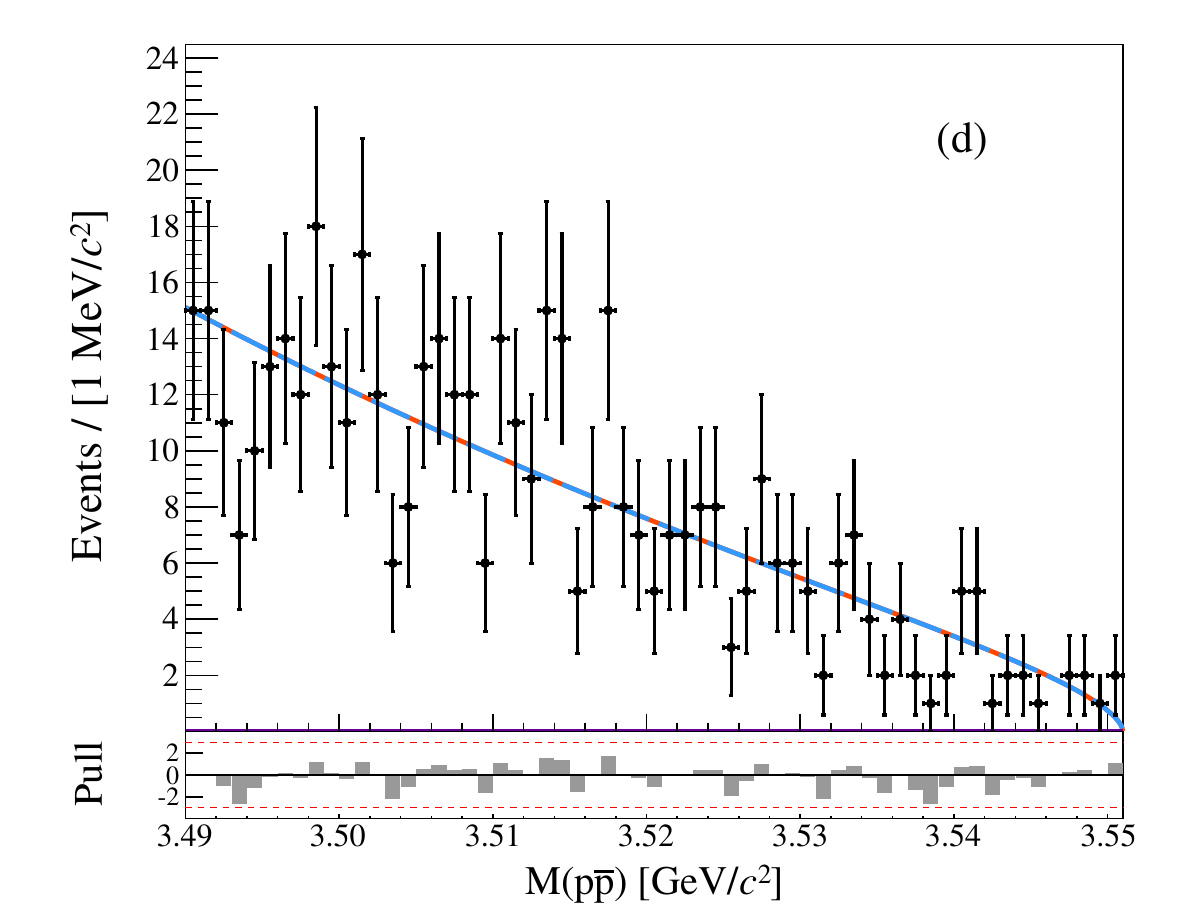}
\end{minipage}

\begin{minipage}[t]{0.445\linewidth}
\includegraphics[width=1\textwidth]{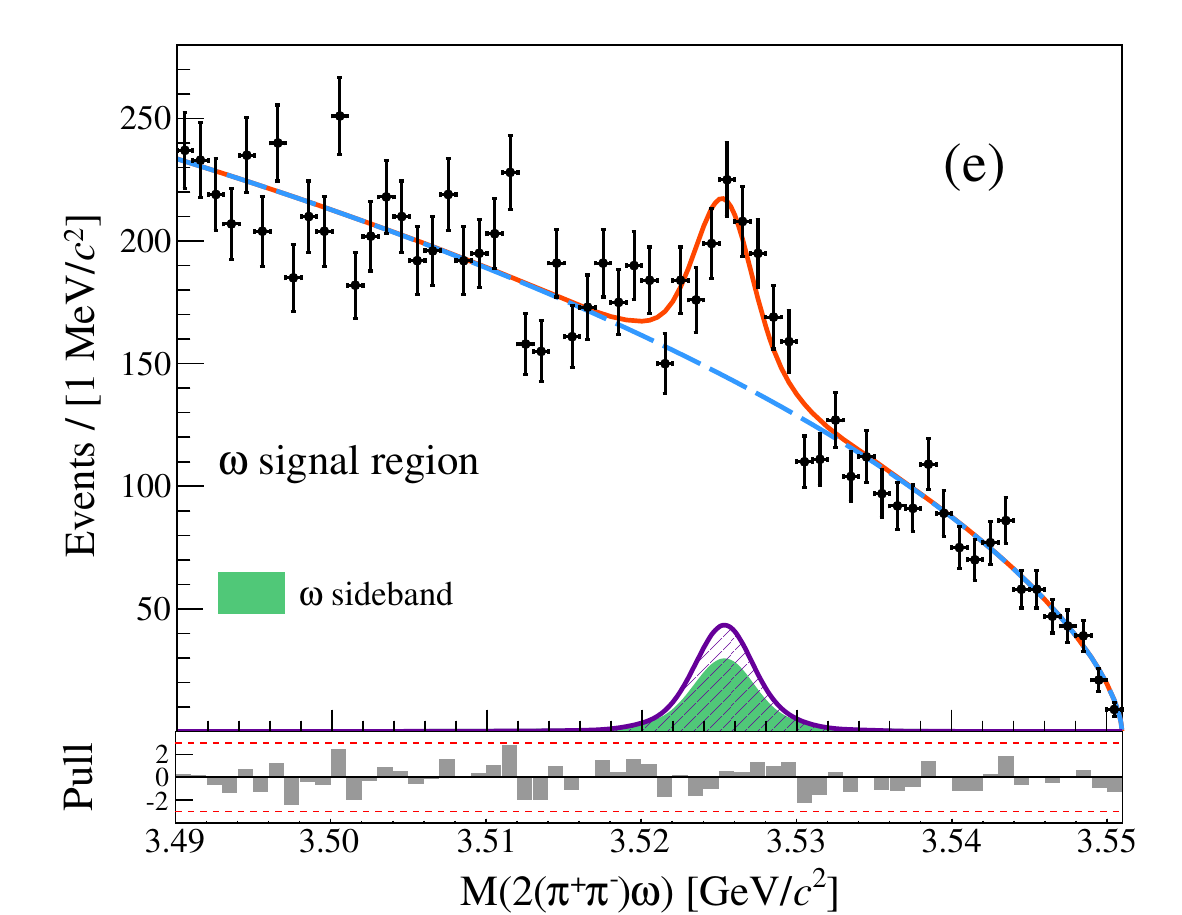}
\end{minipage}
\begin{minipage}[t]{0.445\linewidth}
\includegraphics[width=1\textwidth]{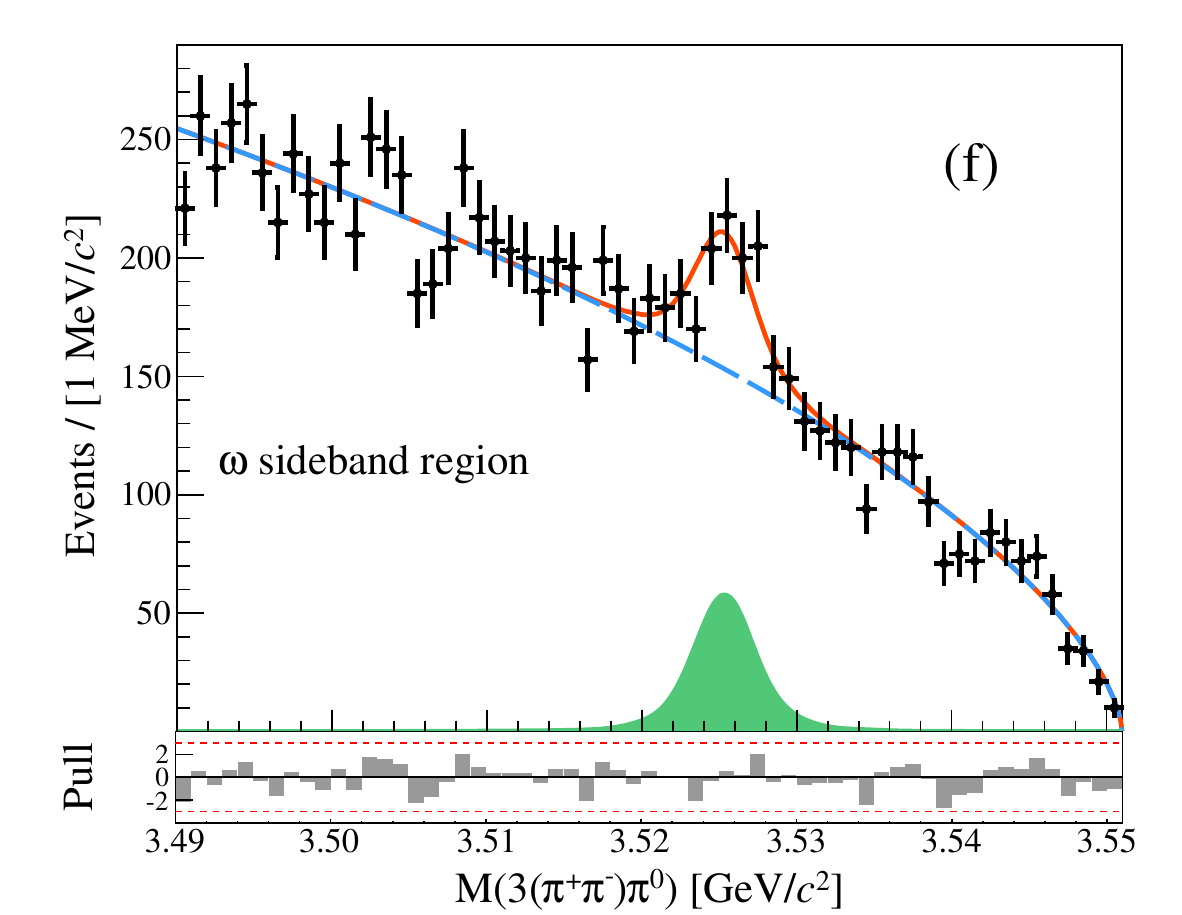}
\end{minipage}

\caption{Fits to the invariant-mass distributions of (a) $h_c\to3(\pi^{+}\pi^{-})\pi^{0}$,
(b) $h_c\to2(\pi^{+}\pi^{-})\pi^{0}\eta$, (c) $h_c\to2(\pi^{+}\pi^{-})\eta$, (d) $h_c\to p\bar{p}$ ,
(e) $h_{c}\to2(\pi^{+}\pi^{-})\omega$ in the $\omega$ signal region and (f) $h_{c}\to3(\pi^{+}\pi^{-})\pi^{0}$ in the $\omega$ sideband region.
The green histogram shows the scaled $h_c$ peak obtained from the fit to the events inside the $\omega$ sideband region.
The lower panels of each plot show the fit residuals expressed in number of standard deviations.}
		\label{fit:five_Mode_mhc}
		
		\end{figure*}

The number of $h_c$ signal events $N_{h_c}^{\rm obs}$ for each signal decay mode is determined from maximum likelihood fits to the corresponding mass spectra, as shown in Fig.~\ref{fit:five_Mode_mhc}.  For the channels where a signal is observed, the signal shape is described by a convolution of the MC-simulated shape and a Gaussian function accounting for the difference in mass resolution between data and MC simulation, where the width of the Gaussian function is a free parameter in the fit.
For the channels with no significant signals observed ($h_{c}\to2(\pi^{+}\pi^{-})\eta$ and $h_{c}\to p\bar{p}$) only the MC-simulated shapes are used. In the fit to the $h_{c}\to2(\pi^{+}\pi^{-})\pi^{0}\eta$ decay, two additional normalized peaking background components from $h_{c}\to\gamma\eta_{c},\eta_{c}\to2(\pi^{+}\pi^{-})\eta$ and $h_{c}\to\gamma\eta_{c},\eta_{c}\to2(\pi^{+}\pi^{-}\pi^{0})$ are included with fixed background contributions of 3.9 and 35.4 events, respectively. Additionally, for the $h_{c}\to2(\pi^{+}\pi^{-})\eta$ mode, a contribution of 20 events is  included from  $h_{c}\to\gamma\eta_{c},\eta_{c}\to\pi^{+}\pi^{-}\eta^{'}$ background.
The remaining non-peaking background shape is described by an ARGUS function~\cite{Argus}, with endpoint fixed at the kinematic threshold of 3.551~GeV/$c^{2}$. The statistical significance of the $h_{c}\to3(\pi^{+}\pi^{-})\pi^{0}$ decay is greater than 5$\sigma$,
and that of the intermediate process $h_{c}\to2(\pi^{+}\pi^{-})\omega$ in the $3(\pp)\pi^0$ final state is 4.7$\sigma$. The significance for the $h_{c}\to2(\pi^{+}\pi^{-})\pi^{0}\eta$ mode is determined to be 4.8$\sigma$. For each signal decay mode, the significance is estimated from the likelihood difference with and without signal component included in the fit, taking into account the change in the number of degrees of freedom.  The systematic uncertainties are considered in determining the final signal significance. The signal yield for $h_c\to 2(\pi^+\pi^-)\omega$ is evaluated by performing a simultaneous fit to the $M(3(\pi^{+}\pi^{-})\pi^{0})$ spectra for the events in both $\omega$ signal and sideband regions using the signal and background shapes described above.
The width of the Gaussian function used for smearing is fixed to the value
obtained from the fit for the $h_{c}\to3(\pi^{+}\pi^{-})\pi^{0}$ decay.
The $h_c$ yield in the $\omega$ sideband region is scaled by a factor $f_\omega = 0.50$, which takes
into account the relative difference in the number of events between $\omega$ signal and sideband regions, as shown in Fig.~\ref{newFig:mass_resolution}. Specifically, in the fitting of the $\pi^{+}\pi^{-}\pi^{0}$ invariant mass spectrum, the signal component is modeled with a double Gaussian function. The remaining combinatorial background shape is described by a first-order polynomial function.

Since no significant signals of $h_{c}\to2(\pi^{+}\pi^{-})\eta$ and $h_{c}\to p\bar{p}$ are observed, the upper limits of the signal yields for these decays are determined to be 33.7 and 12.4, respectively. The upper limit has already incorporated the systematic uncertainty, and the corresponding method is discussed in Section~\ref{sec:sysU}.

The BFs of the $h_c$ decay channels are calculated with
\begin{equation}\label{determine_branch}
\mathcal{B}\left( {{h_c} \to X} \right) = {\textstyle{{{N_{h_{c}}^{\rm obs} }} \over { {N_{\psi \left( {3686} \right) }}\cdot \mathcal{B}\left( {\psi \left( {3686} \right) \to {\pi ^0}{h_c}} \right)\cdot {\prod _{i}}  \mathcal{{B}}_{i} \cdot \varepsilon }}}.
\end{equation}
Here, $N_{h_c}^{\rm obs}$ is the number of observed $h_c$ signal events, $N_{\psi(3686)}$ is the total number of $\psi(3686)$ events, $X$ denotes a given final state, $\Br_{i}$ is the BF of the $i$-th intermediate state taken from the PDG~\cite{pdg2022}, and $\varepsilon$ is the detection efficiency. The results and related numerical information are given in Table~\ref{list_summary}.

\begin {table*}[htbp]
\begin{center}
\begin{small}
    {\caption {The number of observed signal events $N_{h_{c}}^{obs}$, the detection efficiency, the product BF $\Br(\psi(3686)\to\pi^{0}h_{c})\times\Br(h_{c}\to X)$, and the absolute BF $\Br(h_{c}\to X)$. The first and second uncertainties are statistical and systematic, respectively.}
    \label{list_summary}}
    \begin {tabular}{  c |  c | c | c |c | c }\hline\hline

 Final state ($X$) & $N_{h_{c}}^{\rm obs}$ & $\epsilon(\%)$   &   \multicolumn{2}{c|}{$\Br$($\psipp\to\pi^{0}h_c)\times\Br(h_c\to X$)}  & $\Br$($h_c\to X$) \\

   & & &    This study & Previous BESIII result  &  \\    \hline
 	$3(\pi^{+}\pi^{-})\pi^{0}$ & $834.7\pm102.3$ &  $4.64\pm0.01$   &     $\left( {6.79 \pm 0.83 \pm 0.56} \right) \times {10^{ - 6}}$     &   $<7.56\times10^{-6}$~\cite{pdg2022} &    $\left( { 9.28\pm 1.14 \pm 0.77 }  \right) \times {10^{ - 3}}$   \\

	  $2(\pi^{+}\pi^{-})\pi^{0}\eta$   & $154.4\pm30.8$ &  $2.68\pm0.01$   &   $\left( {5.33 \pm 1.10 \pm 0.56 } \right) \times {10^{ - 6}}$   &    --    &  $\left( {7.55 \pm 1.51 \pm 0.77 } \right) \times {10^{ - 3}}$    \\

	$2(\pi^{+}\pi^{-})\eta$   &$<$ 33.7 & $7.05\pm0.01$    &   $<4.53 \times 10^{-7}$    &    --     &   $<6.19 \times 10^{-4}$  \\
	
	$p\bar{p}$    & 12.4 &  $14.35\pm0.02$   &   $<3.22 \times 10^{-8}$    &     $<1.3 \times 10^{-7}$~\cite{lch_prd}  &  $<4.40 \times 10^{-5}$  \\
	
    $2(\pi^{+}\pi^{-})\omega$  & $236.9\pm50.9$ &  $3.42\pm0.01$   &  $\left( {2.93 \pm 0.63 \pm 0.26 } \right) \times {10^{ - 6}}$   &  -- &  $\left( { 4.00 \pm 0.86 \pm 0.35 }  \right) \times {10^{ - 3}}$   \\

\hline
\hline
\end{tabular}
\end{small}

\end{center}
\end{table*}

\section{Systematic Uncertainties}\label{sec:sysU}
The sources of systematic uncertainties in the BF measurements are associated with the  tracking efficiency, photon-detection efficiency, $\pi^{0}$ or $\eta$ reconstruction efficiency, kinematic fit, signal-yield extraction, choice of theoretical model, etc.  The evaluated contributions to the systematic uncertainties are listed in Table~\ref{list_sys} and discussed below.

\begin{itemize}

\item[(i)]{{\bf Tracking efficiency:}}
The tracking efficiency of charged pions is estimated by studying the control sample
of $J/\psi\to\pi^{+}\pi^{-}\pi^{0}$ decays.
The MC simulation is re-weighted in two-dimensional ($\cos\theta, p_{t}$) intervals
according to the efficiencies obtained from the control sample,
where $\theta$ is the polar angle and $p_{t}$ is the transverse momentum of the charged pion.
The observed change in the selection efficiency is assigned as the systematic uncertainty.
The systematic uncertainties associated with  the tracking efficiencies of protons (antiprotons)
are estimated with the control samples $J/\psi \to p\bar{p}\pi^{+}\pi^{-}$ decays~\cite{error_ppb2pi}.
An uncertainty of 1.0\% per track is assigned.

\item[(ii)]{\bf Photon-detection efficiency:} The photon-detection efficiency is studied using a control sample of $e^{+}e^{-}\to\gamma\mu^{+}\mu^{-}$ events. The difference in the photon detection efficiencies between data and MC simulation, 0.5\%, is assigned as the systematic uncertainty for each photon.

\item[(iii)]{\bf $\pi^{0}$ and $\eta$ reconstruction efficiency}: Based on a control sample of $e^{+}e^{-}\to\omega\pi^{0}$ events at $\sqrt{s}=3.773~\rm{GeV}$, the difference of the $\pi^{0}$ reconstruction efficiencies between data and MC simulation is studied and found to have the following dependence on momentum: $(0.06-2.41\cdot p$ [GeV$/c$])\%.
The systematic uncertainty associated with this difference is calculated by averaging
this function over the $\pi^0$ momentum distribution in each signal decay mode.
The uncertainty due to the choice of $\eta$ mass window is determined to be $1.0\%$ per $\eta$, from studies of a high purity control sample of $J/\psi\to p\bar{p}\eta$ decays~\cite{eta_recon}.

\item[(iv)]{\bf Kinematic fit}:
The systematic uncertainty associated with the kinematic fit is estimated by comparing the efficiencies with and without a helix parameter correction applied to the signal MC events~\cite{YPG:bam}. The assigned systematic uncertainties are (2.1-3.2)\% for the different signal decay modes.

\item[(v)]{\bf Signal-yield extraction}:
\item[\textbullet]{\bf Mass window}:
 To evaluate the systematic uncertainty associated with the choice of mass window, we perform a Barlow test~\cite{barlow_test} to examine the deviation in significance ($\zeta$) between the baseline fit and that performed for the systematic test. The deviation in significance is defined as
\begin{equation}
    \zeta=\frac{\left|V_{\text {nominal }}-V_{\text {test }}\right|}{\sqrt{\left|\sigma_{V \text { nominal }}^2-\sigma_{V \text { test }}^2\right|}},
\end{equation}
where $V$ represents the BF; $\sigma_{V}$ is the statistical uncertainty on $V$. For each signal decay, the mass window is varied by up to one times the corresponding mass resolution, with a step of 0.25 times the mass resolution. If the $\zeta$ value is not greater than 2, its effect is assumed to be negligible. For the decays where no significant signal is observed, the corresponding systematic uncertainty is  assigned as the maximum difference  within the various test ranges mentioned earlier (unless otherwise specified, similar situations are estimated using the same method). Additionally, the systematic uncertainty associated with  the choice of $\omega$ mass window is estimated from a control sample of $\psi(3686)\to K^{+}K^{-}\omega$ decays. The differences in the acceptance efficiencies between data and MC simulation  are taken as the corresponding systematic uncertainties.

\item[\textbullet]  {\bf Fit range}:
The systematic uncertainty arising from the choice of  fit range is similarly investigated  using the Barlow test. To determine the corresponding $\zeta$ distribution, we systematically adjust the fit range by shrinking the interval (3480, 3551)~MeV/$c^{2}$ to (3500, 3551)~MeV/$c^{2}$, with a step size of 2~MeV/$c^2$.

\item[\textbullet]  {\bf Signal shape}:
To estimate the systematic uncertainty associated with the choice of signal shape, we use an alternative MC-simulated shape convolved with a double Gaussian function. The difference relative to the nominal fit is taken as the systematic uncertainty.

  \item[\textbullet]  {\bf Background shape}:
The uncertainties due to the choice of background shape is estimated by replacing the ARGUS function with a second-order Chebychev polynomial function. The change in the fitted signal yield is taken as the systematic uncertainty.

 \item[\textbullet]  {\bf $\eta_{c}$ peaking background}:
   The uncertainty associated  with the knowledge of the  $\eta_{c}$ peaking background is estimated according to the uncertainty of the corresponding BF measurements~\cite{pdg2022,new_psi3686Topi0hc_BF} or by including this contribution in the fit, assuming $\Br(\eta_{c}\to2(\pi^{+}\pi^{-})\pi^{0})$ is equal to  $\Br(\eta_{c}\to2(\pi^{+}\pi^{-})\eta)$, and assigning the largest differences between these  and the baseline results as the systematic uncertainty.

\item[\textbullet]  {\bf Wrong-combination background}:
The uncertainty due to the wrong-combination background (WCB) in the $\pi^0$ reconstruction is investigated by studying the matching angle between generated and reconstructed momenta. Events containing candidates with matching angles greater than $\SI{10}{\degree}$ are classified as the WCB background. The possible bias due to WCB is estimated by comparing the signal yields with and without including the WCB contribution in the fits. The difference in the fitted signal yield is taken as the systematic uncertainty.

\item[\textbullet]  {\bf Normalization factor}:
The systematic uncertainty related to the normalization factor in calculating the non-$\omega$ contribution in extracting the number of the $h_{c}\to2(\pi^{+}\pi^{-})\omega$ signal events is estimated by producing multi-dimensional Gaussian random numbers with the covariance matrix from the fit as input. A group of normalization factors is obtained from this procedure and the standard deviation of the corresponding distribution is taken as the systematic uncertainty.

\item[\textbullet]  {\bf\boldmath $\omega$ sideband}:
The systematic uncertainty due to the choice of $\omega$ sideband region is determined by widening and narrowing the range
by one standard deviation of the $\omega$ mass resolution. The largest change relative to the fitted signal yield for $h_c\to 2(\pi^+\pi^-)\omega$ is taken as the corresponding systematic uncertainty.

\item[(vi)]{\bf Theoretical model}:
There is a potential systematic bias associated with the choice of theoretical model for the primary decay of $\psi(3686)\to\pi^{0}h_{c}$ and the sequential decay $h_{c}\to \text{hadrons}$, since these decays are not understood well at present. The difference in reconstruction efficiencies between simulated events generated with the PHSP model and the model proposed in Refs.~\cite{evtgen_a,evtgen_b} is assigned as the systematic uncertainty for $\psi(3686)\to \pi^0h_c$. The difference in efficiencies between events generated with the PHSP model and the mixed model including additional intermediate states is taken as the systematic uncertainty for $h_c\to$ hadrons.

\item[(vii)]{\bf MC sample size}:
The statistical uncertainty in the reconstruction efficiency associated with the finite MC sample size can be
calculated as $\Delta_{\epsilon}=\sqrt{\epsilon(1-\epsilon)/N}$, where $\epsilon$ is the reconstruction efficiency
and $N$ is the total number of generated events. The relative uncertainty from this source  is $\Delta_{\epsilon}/\epsilon$.

\item[(viii)]{\bf Input BFs}:
The uncertainties in the knowledge of the BFs of $\Br(\psi(3686)\to\pi^{0}h_{c})$, $\Br(\eta\to\gamma\gamma)$, and $\Br(\omega\to\pi^{+}\pi^{-}\pi^{0})$, which are 5.6\%~\cite{new_psi3686Topi0hc_BF},
0.5\%~\cite{pdg2022}, and 0.8\%~\cite{pdg2022}, respectively, are assigned as individual systematic uncertainties.

\item[(ix)]{\boldmath $N_{\psi(3686)}$}:
The total number of $\psi(3686)$ events in the sample is determined from inclusive hadronic $\psi(3686)$ decays with an uncertainty of 0.5\%~\cite{psip_num_0912}.
\end{itemize}

The systematic uncertainties are summarized in Table~\ref{list_sys}. For each signal decay mode
the total systematic uncertainty is obtained by adding all contributions in quadrature under the assumption that they are independent.

\begin{figure*}[htbp]

\begin{minipage}[t]{0.445\linewidth}
\includegraphics[width=1\textwidth]{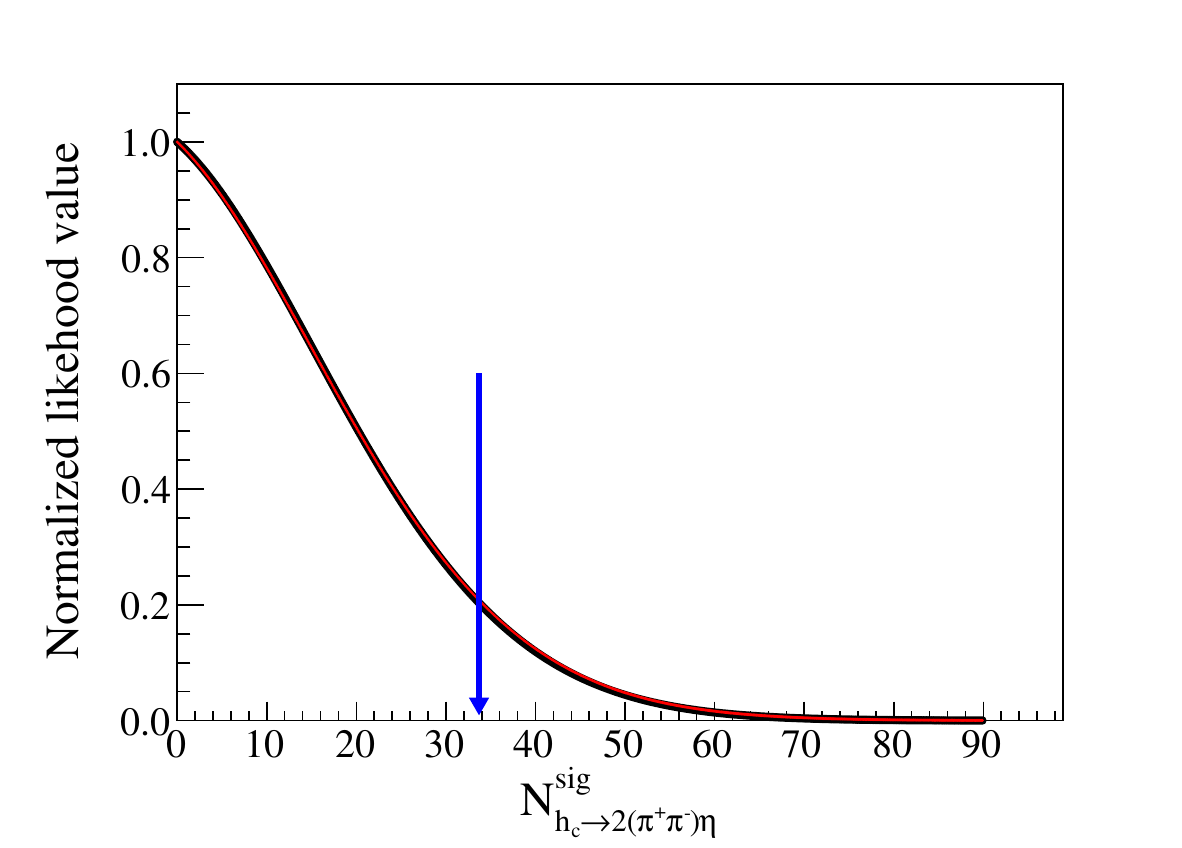}
\end{minipage}
\begin{minipage}[t]{0.445\linewidth}
\includegraphics[width=1\textwidth]{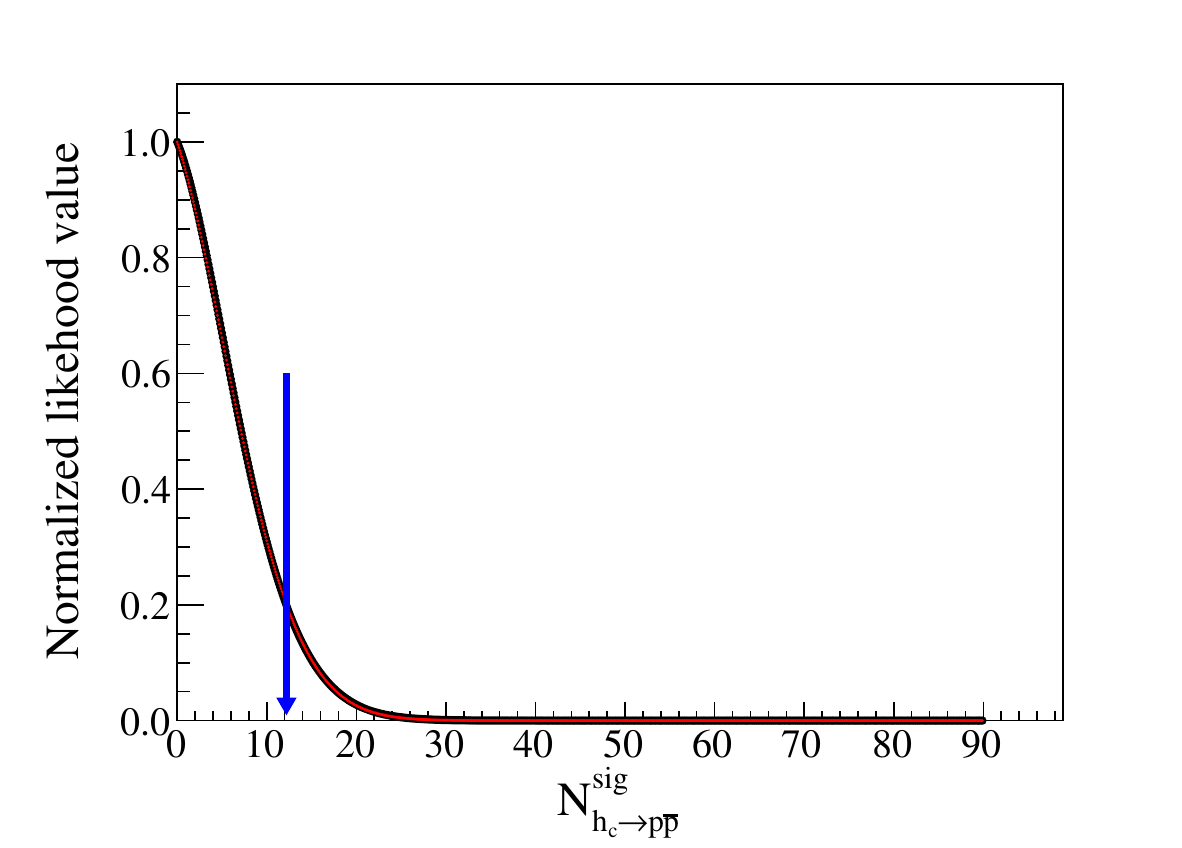}
\end{minipage}

\caption{The normalized likelihood distribution for (up panel) $h_{c}\to2(\pi^{+}\pi^{-})\eta$ and (down panel) $h_{c}\to p\bar{p}$ mode. The results
obtained with and without incorporating the systematic uncertainties are shown in blue dots and black dots, respectively. The
arrow is the position of the upper limit on the signal yields at
90\% confidence level.}
\label{fig:upperlimits}
\end{figure*}

 The systematic uncertainty from the upper limit of signal events $N_{h_{c}}^{\rm obs}$ at 90\% confidence level includes additive sources and multipicative sources. To account for the additive systematic uncertainties related to the fits, several alternative fits are performed. These alternative fits involve signal shape, background shape, $\eta_{c}$ peaking-background shape as well as different fit ranges and the one resulting the most conservative upper limit is chosen. Then, the multiplicative systematic uncertainty is incorporated in the calculation of the upper limit via~\cite{Stenson:2006gwf, Liu:2015uha}
\begin{equation}
   L'(N) = \int_0^1 {L({\textstyle{S \over {\hat S}}}N)} \exp \left[ { - {\textstyle{{(S - \hat S)} \over {2\sigma _S^2}}}} \right]dS,
\end{equation}
where, $L(N)$ is the likelihood distribution as a function of signal events, $N$; $S$  is the expected efficiency and $\hat S$ is the nominal efficiency; $\sigma_{S}$ is its multiplicative systematic uncertainty coming from Table~\ref{list_sys}. Following these steps, the corresponding likelihood distribution can be obtained as shown in Fig.~\ref{fig:upperlimits}. The assumed yields for processes $h_{c}\to2(\pi^{+}\pi^{-})\eta$ and $h_{c}\to p\bar{p}$ at 90\% confidence level are set as upper limits, and the corresponding BF calculated using Equation~\ref{determine_branch} is the upper limit BF, as shown in Table ~\ref{list_summary}.

\begin {table*}[htbp]
\centering
    {\caption {Relative systematic uncertainties (in percent) for each $h_c$ decay channel. The dash indicates that the systematic uncertainty is not applicable.  Asterisk denotes that the additive systematic uncertainty values are invalid.}
    \label{list_sys}}

    \begin {tabular}{l | c | c | c | c | c}\hline\hline

Source & $3(\pi^{+}\pi^{-})\pi^{0}$ & $2(\pi^{+}\pi^{-})\pi^{0}\eta$ & $2(\pi^{+}\pi^{-})\eta$  & $\ppb$   &   $2(\pi^{+}\pi^{-})\omega$   \\   \hline
	 Tracking                       &   2.3        & 1.6          & 1.4            & 2.0               & 2.3  \\
	Photon efficiency               &   2.0        & 3.0          & 2.0            & 1.0               & 2.0      \\
	$\pi^{0}$ reconstruction        &   2.1        & 2.2          & 0.5            & 0.5               & 1.6   \\
	$\eta$ reconstruction           &   --         & 1.0          & 1.0            & --                & -- \\
	Kinematic fit                   &   2.1        & 2.3          & 2.5            & 3.2               & 2.3     \\
	Mass window                     &   negligible & 1.2          & 6.8            & --                & 1.1 \\
    Signal shape	                &   0.8        & 0.3          &   *            &   *               & 2.1\\
    Background shape                &   2.6        & 4.1          &   *            &   *               & 2.5 \\
$\eta_{c}$ peaking-background shape &   --         & 1.8          &   *            & --                & --\\
    WCB                             &   1.3        & 1.9          & --             & --                & negligible\\
    Fit range	                    &   negligible &  negligible  &                &                   & negligible \\
   Normalization factor           	&   --         & --           & --             & --                & 0.5  \\
  $\omega$ sideband	                &   --         & --           & --             & --                & 3.8 \\

    Theoretical model               &   3.1        &  5.0         & 0.1            & negligible        &  0.6\\

   MC sample size                   &   0.2        & 0.1          &  0.1           & 0.3               & 0.4 \\

   Input BFs                        &   5.6        &  5.6         &  5.6           & 5.6               & 5.7\\

   $N_{\psi(3686)}$                 &   0.5        & 0.5          & 0.5            & 0.5               & 0.5    \\ \hline
	Sum                             &   8.3        & 10.2         & 9.6           & 6.9            & 8.8   \\

\hline
\hline
\end{tabular}

\end{table*}

\section{Summary}
By analyzing $(2712.4\pm14.1)\times10^{6}$ $\psi(3686)$ events collected in the BESIII experiment, we report the observation of $h_{c}\to3(\pi^{+}\pi^{-})\pi^{0}$, as well as evidence for the decays $h_{c}\to2(\pi^{+}\pi^{-})\pi^{0}\eta$ and $h_c\to2(\pi^{+}\pi^{-})\omega$. We have also searched for $h_{c}\to2(\pi^{+}\pi^{-})\eta$ and $h_{c}\to p\bar{p}$, but no significant signal is observed for either decay. The measured BFs or upper limits at the 90\% confidence level are listed in Table~\ref{list_summary}. The upper limit on $\Br(h_{c}\to p\bar{p})$  is a factor of three lower than that set in  previous studies~\cite{lch_prd}. Furthermore, the limit is a  factor of 100 lower than the predictions found in Refs.~\cite{lch_ref5,lch_ref6}, which indicates the need for improved  theoretical calculations as well as more sensitive measurements.

	\input{acknowledgement.tex}


\end{document}

%% file: def-com-new.tex
\definecolor{boslv}{rgb}{0.0, 0.65, 0.58}
\definecolor{Munsell}{HTML}{00A877}

\newcommand{\psipp}{\psi(3686)}

\newcommand{\Br}{\mathcal{B}}

\newcommand{\ppb}{p\bar{p}}

\newcommand{\pp}{\pi^+\pi^-}

\newcommand{\bfg}{\begin{figure}}
\newcommand{\efg}{\end{figure}}
\newcommand{\bitm}{\begin{itemize}}
\newcommand{\eitm}{\end{itemize}}
\newcommand{\bnum}{\begin{enumerate}}
\newcommand{\enum}{\end{enumerate}}
\newcommand{\btbl}{\begin{table}}
\newcommand{\etbl}{\end{table}}
\newcommand{\btbu}{\begin{tabular}}
\newcommand{\etbu}{\end{tabular}}
\newcommand{\bcl}{\begin{center}}
\newcommand{\ecl}{\end{center}}

\newcommand{\beq}{\begin{equation}}
\newcommand{\eeq}{\end{equation}}
\newcommand{\beqr}{\begin{eqnarray}}
\newcommand{\eeqr}{\end{eqnarray}}

%% file: authorlist_2023-11-13.tex
\author{
\begin{small}
\begin{center}
M.~Ablikim$^{1}$, M.~N.~Achasov$^{4,c}$, P.~Adlarson$^{75}$, O.~Afedulidis$^{3}$, X.~C.~Ai$^{80}$, R.~Aliberti$^{35}$, A.~Amoroso$^{74A,74C}$, Q.~An$^{71,58,a}$, Y.~Bai$^{57}$, O.~Bakina$^{36}$
% ...
\\
\vspace{0.2cm}
(BESIII Collaboration)\\
\vspace{0.2cm} {\it
$^{1}$ Institute of High Energy Physics, Beijing 100049, People's Republic of China\\
% ...
$^{3}$ Bochum  Ruhr-University, D-44780 Bochum, Germany\\
$^{4}$ Budker Institute of Nuclear Physics SB RAS (BINP), Novosibirsk 630090, Russia\\
% ...
$^{35}$ Johannes Gutenberg University of Mainz, Johann-Joachim-Becher-Weg 45, D-55099 Mainz, Germany\\
$^{36}$ Joint Institute for Nuclear Research, 141980 Dubna, Moscow region, Russia\\
% ...
$^{57}$ Southeast University, Nanjing 211100, People's Republic of China\\
$^{58}$ State Key Laboratory of Particle Detection and Electronics, Beijing 100049, Hefei 230026, People's Republic of China\\
% ...
$^{71}$ University of Science and Technology of China, Hefei 230026, People's Republic of China\\
% ...
$^{74}$ University of Turin and INFN, (A)University of Turin, I-10125, Turin, Italy; (B)University of Eastern Piedmont, I-15121, Alessandria, Italy; (C)INFN, I-10125, Turin, Italy\\
$^{75}$ Uppsala University, Box 516, SE-75120 Uppsala, Sweden\\
% ...
$^{80}$ Zhengzhou University, Zhengzhou 450001, People's Republic of China\\
\vspace{0.2cm}
$^{a}$ Deceased\\
% ...
$^{c}$ Also at the Novosibirsk State University, Novosibirsk, 630090, Russia\\
% ...
}\end{center}
\vspace{0.4cm}
\end{small}
}

%% file: acknowledgement.tex
\acknowledgements
The BESIII Collaboration thanks the staff of BEPCII and the IHEP computing center for their strong support. This work is supported in part by National Key R\&D Program of China under Contracts Nos. 2020YFA0406300, 2020YFA0406400; National Natural Science Foundation of China (NSFC) under Contracts Nos. 11635010, 11735014, 11835012, 11935015, 11935016, 11935018, 11961141012, 12022510, 12025502, 12035009, 12035013, 12061131003, 12192260, 12192261, 12192262, 12192263, 12192264, 12192265, 12221005, 12225509, 12235017, 12150004; Program of Science and Technology Development Plan of Jilin Province of China under Contract No. 20210508047RQ and 20230101021JC; the Chinese Academy of Sciences (CAS) Large-Scale Scientific Facility Program; the CAS Center for Excellence in Particle Physics (CCEPP); Joint Large-Scale Scientific Facility Funds of the NSFC and CAS under Contract No. U1832207; CAS Key Research Program of Frontier Sciences under Contracts Nos. QYZDJ-SSW-SLH003, QYZDJ-SSW-SLH040; 100 Talents Program of CAS; The Institute of Nuclear and Particle Physics (INPAC) and Shanghai Key Laboratory for Particle Physics and Cosmology; European Union's Horizon 2020 research and innovation programme under Marie Sklodowska-Curie grant agreement under Contract No. 894790; German Research Foundation DFG under Contracts Nos. 455635585, Collaborative Research Center CRC 1044, FOR5327, GRK 2149; Istituto Nazionale di Fisica Nucleare, Italy; Ministry of Development of Turkey under Contract No. DPT2006K-120470; National Research Foundation of Korea under Contract No. NRF-2022R1A2C1092335; National Science and Technology fund of Mongolia; National Science Research and Innovation Fund (NSRF) via the Program Management Unit for Human Resources \& Institutional Development, Research and Innovation of Thailand under Contract No. B16F640076; Polish National Science Centre under Contract No. 2019/35/O/ST2/02907; The Swedish Research Council; U. S. Department of Energy under Contract No. DE-FG02-05ER41374.